\documentclass[preprint,aps,onecolumn,showpacs,superscriptaddress,amsmath,amssymb]{revtex4-1}
\usepackage{graphicx}
\usepackage{amsmath}
\usepackage{amsfonts}
\usepackage{amssymb}
\usepackage{graphicx}
\usepackage{epstopdf}
\usepackage[colorlinks={true}]{hyperref}
\hypersetup{citecolor={blue}, filecolor={blue}, linkcolor={blue}, urlcolor={blue}}
\usepackage{multirow}
\usepackage{mathrsfs}
\usepackage{cases}
\usepackage{txfonts}
\usepackage{color}
\begin{document}
\preprint{}
\title{Complete elimination  of  nonlinear light-matter interactions with broadband ultrafast laser pulses}
\author{Chuan-Cun Shu}\email{c.shu@unsw.edu.au}
\affiliation{School of Engineering and Information Technology,
University of New South Wales, Canberra,  ACT 2600, Australia}
\affiliation{Department of Chemistry, Technical University of Denmark, Building 207, DK-2800 Kongens Lyngby, Denmark}
\author{Daoyi Dong}
\affiliation{School of Engineering and Information Technology,
University of New South Wales, Canberra,  ACT 2600, Australia}
\affiliation{Department of Chemistry, Princeton University, Princeton, New Jersey 08544, USA}
\author{Ian R. Petersen}
\affiliation{School of Engineering and Information Technology,
University of New South Wales, Canberra, ACT 2600, Australia}
\author{Niels E. Henriksen}\email{neh@kemi.dtu.dk}
\affiliation{Department of Chemistry, Technical University of Denmark, Building 207, DK-2800 Kongens Lyngby, Denmark}
\begin{abstract}
The absorption  of a single photon that excites a quantum system from a low to a high energy level is an elementary process of light-matter interaction, and a route towards realizing pure single-photon absorption has both fundamental and practical implications in quantum technology.
Due to nonlinear optical effects, however, the probability of pure single-photon absorption is usually very low, which is particularly pertinent in the case of strong ultrafast laser pulses with broad bandwidth. Here we demonstrate theoretically a counterintuitive coherent single-photon absorption scheme by eliminating nonlinear interactions of ultrafast laser pulses with quantum systems. That is, a completely linear response of the system with respect to the spectral energy density of the incident light at the transition frequency can be obtained for all transition probabilities between 0 and 100\% in a multi-level quantum systems. To that end, a new multi-objective optimization algorithm  is developed to find an optimal  spectral phase of an ultrafast laser pulse, which is capable of eliminating all possible nonlinear optical responses while maximizing the probability of single-photon absorption between quantum states. This work not only deepens our understanding of light-matter interactions, but also offers a new way to study  photophysical and photochemical processes in the ``absence" of nonlinear optical effects.
\end{abstract}
\pacs{32.80.Qk,33.80.-b,42.50.Hz,42.65.-k}%
\maketitle
Exploring  the interaction of light with matter (i.e., atoms and molecules)
at the ultimate limit of single photons is a
topic of much current interest in many disciplines of  science.
This includes topics as, generating single photon sources \cite{NPh:10:19,prl:116:023602,nc:7:10853},
storing single photons in quantum memory devices \cite{phys.today:qmp,prl:114:053602},
and controlling the interactions between single photons and matter \cite{prl:113:053601,prl:108:093601,prl:117:043601}.
When a beam of light interacts with matter with quantized  energy levels,
optical absorption and emission are fundamental processes corresponding to a transition from one energy
level to another. The rate of absorption has
a component proportional to the energy density of the beam.  The transition
rate also contains terms of higher order, i.e. nonlinear terms in the
energy density. The linear term in the absorption rate corresponds to the excitation in which a single photon is absorbed,
whereas the nonlinear terms correspond to the excitation in which two or more photons are absorbed \cite{loudon}.\\ \indent
 The probability of pure single-photon absorption under normal circumstances is very low. One of the major difficulties in realizing single-photon absorption with unit
probability is to overcome decoherence (e.g., population relaxation) due to  the intrinsic fragility of quantum states, which is also a common challenge for quantum technology.
Ultrafast laser pulses  provide an alternative approach to manipulate many quantum processes on extremely short time
scales (atto- to picoseconds) before decoherence plays a role \cite{arpc:60:277,science:350:790,nc:7:11200,nc:8:238}. When such a laser pulse that contains a huge number of photons
within the broad bandwidth excites  matter, another major difficulty due to nonlinear optical effects may emerge.
Based on a mature spectral phase-shaping technique \cite{rsi:98:1929}, considerable theoretical and experimental effort has been directed toward
the study of single-photon phase control in the weak-field regime
\cite{PNAS:109:19578,science:313:1257,jpcl:6:4032,njp:12:015003,jpcl:6:824,jcp:134:164308}. Related theoretical work has shown that a {\emph{nearly}} linear response of the system as a function of
laser energy density is possible, somewhat beyond the strictly weak-field limit,
provided appropriate laser spectral-phase modulation is introduced \cite{prl:108:183002}.
Furthermore, seminal work by Silberberg \emph{et al.} \cite{arpc:60:277,nature:396:239,prl:86:47} has demonstrated how
this approach can be used in the
modulation  of multi (two)-photon transitions in atoms, and a direct signature has been observed in which a two-photon nonlinear optical process in molecules can be significantly affected by chirping the spectral phase of the laser pulse at very low light intensities \cite{jpcl:3:2458}.  Inspired by these previous studies, a fundamentally important but largely unexplored  question is whether
pure spectral phase shaping of an ultrafast  laser pulse  can lead to
complete elimination of nonlinear optical effects.    \\ \indent
 In this Letter, we demonstrate that a completely linear absorption
probability  - as a function of the energy density at the transition frequency - can be obtained
for {\em all} transition probabilities   between the minimum 0 and the maximum 100\% in a prototypical multi-level quantum system.
To that end,
we develop a monotonically convergent multi-objective optimization algorithm, which combined with a perturbation theory analysis is utilized to find the optimal spectral phase of a laser pulse for  minimizing   nonlinear optical effects while maximizing the probability of single-photon transition. The robustness of the maximal single-photon absorption against the influence of spectral field noise is  examined. \\ \indent
\begin{figure}[!t]\centering
\resizebox{0.5\textwidth}{!}{%
  \includegraphics{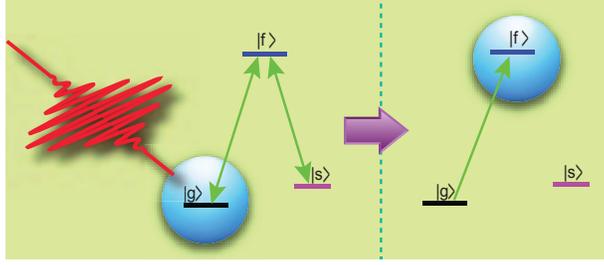}
} \caption{The interaction of a laser field with a  three-level
($\Lambda$-type) quantum system, consisting of two lower levels $|g\rangle$ and $|s\rangle$ and an upper level $|f\rangle$.
A broadband ultrafast laser pulse is  shaped to obtain a linear response of the population in the state $|f\rangle$  with respect to the energy density  while eliminating all possible transitions to state $|s\rangle$.}  \label{fig1}
\end{figure}
The basic aim of this work is sketched in Fig. \ref{fig1}. We consider the simplest multi-level
quantum mechanical system (atom or molecule),  consisting of two lower states $|g\rangle$ and $|s\rangle$
and an excited state $|f\rangle$ with eigenenergies $E_g<E_{s}<E_f$.
The transitions between the lower states
$|g\rangle$, $|s\rangle$ and the excited states $|f\rangle$ are dipole allowed, but the transition between the
states $|g\rangle$ and $|s\rangle$ is dipole forbidden.
An ultrafast laser pulse is used to excite such a
quantum system, whose lifetimes in the states $|s\rangle$ and $|f\rangle$ are
assumed to be much longer than the duration of the laser pulse.
Such a multi-level model has been used for modeling a variety of optical schemes,
including electromagnetically induced transparency (EIT) \cite{jmo:45:471,prl:111:033601}
and stimulated Raman adiabatic passage (STIRAP) \cite{arpc:52:763,jcp:142:170901,prl:99:173001},
where nonlinear multi (two)-photon transitions to the state $|s\rangle$ are taken advantage of
in cancellation of population in the  state $|f\rangle$.
In contrast with these schemes, the present work aims  to protect the absorption from
$|g\rangle$ to $|f\rangle$ via destructive quantum
interferences with multi-photon transition pathways.\\ \indent
The total Hamiltonian operator $\hat{H}(t)$ of the quantum system in interaction with a light field $\mathcal{E}(t)$ can be described  by  $\hat{H}(t)=\hat{H}_0-\hat{\mu}\mathcal{E}(t)$, where $\hat{H}_0$ is the field-free Hamiltonian operator and $\hat{\mu}$ the dipole operator. The wave function $|\Psi(t)\rangle$ of the quantum system, initially in the ground state $|g\rangle$, can be iteratively expanded to arbitrary order as
$|\Psi(t)\rangle=\sum_{k=0}^{\infty}|\psi^{(k)}(t)\rangle$
with $|\psi^{(0)}(t)\rangle=|g\rangle$ and $|\psi^{(k+1)}(t)\rangle=-i\int_{-\infty}^tdt'\mathcal{E}(t')\hat{\mu}_I(t')|\psi^{(k)}(t')\rangle,$
where $\hat{\mu}_I(t)=\exp(i\hat{H}_0t)\hat{\mu}\exp(-i\hat{H}_0t)$ is the dipole operator in the interaction picture.
The electric field of the laser pulse can be expressed as
  \begin{eqnarray}
\mathcal{E}(t)=\frac{1}{2\pi}\mathrm{Re}\left[\int_{0}^{\infty}\boldsymbol{E}(\omega)\exp(-i\omega t)d\omega\right], \label{phase-only}
\end{eqnarray}
with the complex-valued spectral field $\boldsymbol{E}(\omega)=A(\omega)\exp[i\phi(\omega)]$  in terms of the real-valued spectral amplitude $A(\omega)\geq0$ and the real-valued spectral  phase $\phi(\omega)$.
The energy of such a pulse can be expressed as $\mathbb{E}_p\propto\int_{-\infty}^{\infty}\mathcal{E}^2(t) dt\propto \int_{0}^{\infty} A^2(\omega) d\omega$, which is independent of the spectral phase $\phi(\omega)$.
To first order,
the transition probability from the ground state $|g\rangle$ to the final state $|f\rangle$
- corresponding to single-photon absorption -  is given by
\begin{eqnarray}
P^{(1)}_{f} &=& |\langle f|\psi^{(1)}(\infty)\rangle|^2=\mu_{fg}^2 A^2(\omega_{fg}),
\label{SPA}
\end{eqnarray}
where $\mu_{fg}=\langle f|\mu|g\rangle$ is the transition dipole moment and $\omega_{fg}=E_f-E_g$ is the transition frequency between the  states $|g\rangle$ and $|f\rangle$.
Thus, the probability of absorption
depends linearly on the square of the spectral amplitude $A^2(\omega)$ at the resonant transition frequency
$\omega_{fg}$, i.e. the spectral energy density,
but  is independent of the spectral phase.
Furthermore, beyond first order in the interaction,
 odd-order perturbation terms  will contribute to the
transition probability to the state $|f\rangle$ and even-order terms will transfer population to  the states $|s\rangle$ and $|g\rangle$, and a dependence on the spectral phase $\phi(\omega)$ is observed.
The present work will show a coherent control scheme  to realize an
interesting limit of linear absorption from $|g\rangle$ to
$|f\rangle$ by modulating the spectral phase of a
single ultrashort pulse, where the effects of  all higher-order perturbation terms  are eliminated.  \\ \indent
Solutions to analytically unaccessible  maximization/mimization problems under consideration can be established in the framework of quantum optimal control theory (QOCT) \cite{ejpd,jpb:4:R175,njp:12:075008,prl:89:157901,pra:93:053418,pra:93:033417}. Due to the technical complexity
involved in acquiring either monotonic convergence
or general applicability of the algorithms, however,
the present problem is a challenge to the previously developed QOCT methods.  We develop here a gradient-based multi-objective optimization algorithm that not only is capable of  optimizing the spectral phase  of the laser pulse in the frequency domain, but also ensures monotonic  convergence to \emph{each} control objective simultaneously.
{To formulate this method in an elegant mathematical  form, a dummy variable $x\geq0$  is employed  to parameterize  the spectral phase $\phi(\omega)$  with $\phi(x, \omega)$.
As $x$ increases, the change of the final population $P_\ell=|\langle\ell|\Psi(\infty)\rangle|^2$ in an arbitrary quantum  state $|\ell\rangle$ of the system can be written using the chain rule as
 \begin{equation}\label{obj}
\frac{d P_{\ell}}{d x}=\int_{0}^{\infty}\frac{\delta
P_\ell}{\delta \phi(x,\omega)}\frac{\partial\phi(x,\omega)}{\partial
x}d\omega.
\end{equation}
The spectral phase is updated from $\phi(x, \omega)$ to $\phi(x+dx, \omega)$ with
\begin{eqnarray}
\frac{\partial \phi(x,\omega)}{\partial
x}=\int_{0}^{\infty}S(\omega'-\omega)\sum_{\ell, \ell'=1}^M
k_{\ell}(x)\left[\Gamma^{-1}\right]_{\ell\ell'}
\frac{\delta
P_{\ell'}}{\delta \phi(x,\omega')}d\omega',
\label{control
field1}
\end{eqnarray}
 where the convolution function $S(\omega'-\omega)$ is a filter for smoothing the updated spectral phase, and $\Gamma$ is a symmetric matrix composed of the elements $\Gamma_{\ell\ell'}=\int_{0}^{\infty}\delta
P_\ell/\delta \phi(x,\omega)\int_{0}^{\infty}S(\omega'-\omega)\delta
P_{\ell'}/\delta \phi(x,\omega')d\omega'd\omega$.  By inserting Eq.\,(\ref{control field1}) into  Eq.\,(\ref{obj}) (see details in Supplemental Material), we can verify that $P_\ell$ can be monotonically increased (decreased) simultaneously  with $k_\ell>0$ ($k_\ell<0$). Note that the optimization algorithm indicated in Eq. (\ref{control
field1}) is independent of the dimension of Hamiltonian, ensuring its applicability to complex multi-level quantum systems.  To perform this algorithm, the quantum system is driven with an initial guess of the spectral phase $\phi(x_0,\omega)=0$ associated with the temporal field $\mathcal{E}(x_0,t)$, and the generated wavefunction $\Psi(t)$ is used to calculate the gradients of $P_\ell$ with respect to the spectral phase $\phi(x_0,\omega)$  for getting the first gradient $\partial\phi(x_0,\omega)/\partial x$ (see details in Supplemental Material). Equation
(\ref{control field1}) is solved (e.g., by using the Euler method) to obtain a new spectral phase $\phi(x_1=x_0+dx,\omega)$ and the corresponding time-dependent laser field $\mathcal{E}(x_1,t)$ is calculated according to
Eq.\,(\ref{phase-only}).
The spectral phase is iteratively updated from $\phi(x_1,\omega)$ to $\phi(x_2=x_1+dx,\omega), \cdots, \phi(x_n,\omega)$ until
$P_\ell$ converges to the desired control objectives.\\ \indent
To eliminate the nonlinear light-matter interaction terms,  we employ this optimization algorithm to optimize $P_f$ to be as close to the linear absorption probability $P^{(1)}_f$ as possible  while minimizing $P_s$ in the state $|s\rangle$. The unshaped laser field $\mathcal{E}(x_0,t)$ is taken to be  an experimentally accessible Gaussian  transform limited  pulse  with the center frequency of $\omega_0=12500$ cm$^{-1}$ (800 nm) and the full-width at half-maximum (FWHM) of $30$ fs. The eigenenergies of the three-level quantum system are chosen as $E_g=0$, $E_s=0.02\omega_0$, $E_f=\omega_0$, and the transition dipole moments between the two lower states and the excited state are set to $\mu_{gf}=\mu_{sf}=1.0$ a.u. for convenience. A normalized Gaussian spectral filter $S(\omega'-\omega)=\exp[-\mathrm{4ln}2(\omega'-\omega)^2/\sigma^2]$ with a bandwidth of $\sigma=80$ cm$^{-1}$ is used in Eq. (\ref{control field1}). Figure \ref{fig2}(a) shows the final populations in the three states as a function of $A^2(\omega_0)$ with constant
$\phi(\omega) = 0$ spectral phase. The linear optical transition to the state $|f\rangle$, observed in the weak-field limit regime, is significantly affected as the energy density increases, resulting in population transfer to the state $|s\rangle$.
\\ \indent
\begin{figure}[!t]\centering
\resizebox{0.5\textwidth}{!}{%
  \includegraphics{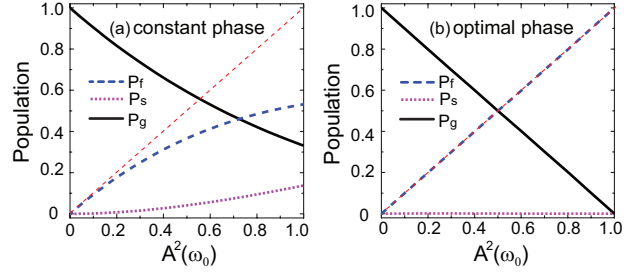}
} \caption{The final populations in the three states with (a) constant spectral phase, (b) optimal spectral phases with respect to $A^2(\omega_0)$,
proportional to the pulse energy.
The dashed line shows the linear scaling of the
transition probability $P_f$ to the upper level, which is valid at
low transition probabilities with a constant laser phase and for all
transition probabilities with an optimized phase.}\label{fig2}
\end{figure}
\begin{figure}[!t]\centering
\resizebox{0.5\textwidth}{!}{%
  \includegraphics{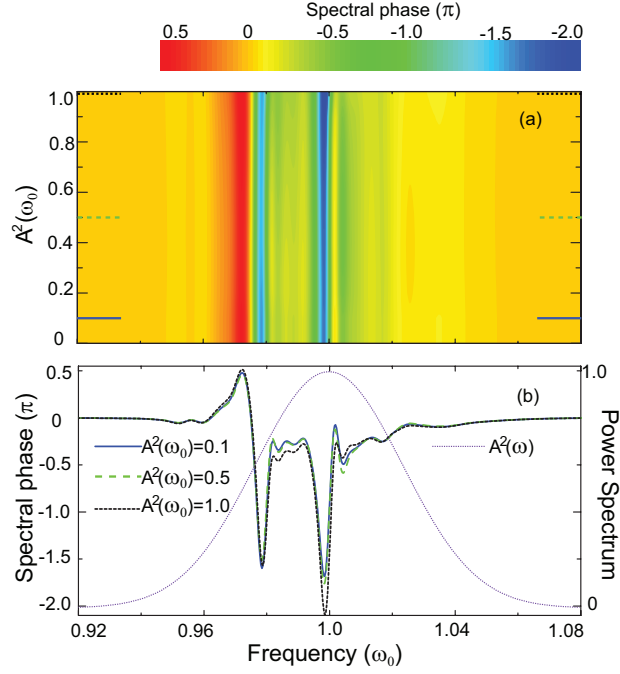}
} \caption{The optimized spectral phases used in Fig. \ref{fig2} (b). (a) The optimized spectral phase as a function of $A^2(\omega_0)$. (b) The optimized spectral phases of the laser pulse with $A^2(\omega_0)=0.1$ (blue line), $0.5$ (green line) and $1.0$ (black line). The dashed line shows
(except for the overall scaling $A^2(\omega_0)$) the
normalized power spectrum, i.e.
the fixed spectral distribution of the laser pulses.}
\label{fig3}
\end{figure}
As seen from Eq.\,(\ref{SPA}), choosing the square of the (peak) spectral amplitude at the critical value of
$A^2(\omega_0)=1.0$ corresponding to $P^{(1)}_f=1.0$,
could in the absence of nonlinear interactions, completely excite the quantum system from the ground state $|g\rangle$ to the final state $|f\rangle$.
We firstly fix the spectral amplitude at $A(\omega_0)=1.0$ (corresponding to the peak intensity of  $I_0=8.0544\times10^{10}$ W/cm$^2$ for the transform limited  pulse),
and then use the optimization algorithm to maximize $P_f$ ($k_f>0$) while minimizing $P_s$ ($k_s<0$). Our results show that by iteratively optimizing the spectral phase,
$P_f$ can be maximized to unity with a high precision (see Fig. 1 in Supplemental Material, where $P_f>0.99999$ and $P_s<1.0\times10^{-8}$). Furthermore,
by using this optimized spectral phase as the initial input,  we further examine the final population responses of the three states with respect to the laser pulses with $A^2(\omega_0)<1.0$ (see Fig. 2 in Supplemental Material).
The nonlinear optical transitions to the intermediate state $|s\rangle$ can be greatly reduced to $P_s<5\times10^{-3}$ and the final population $P_f$ is always greater
than $P^{(1)}_f$ for all of $A^2(\omega_0)<1.0$.
This result provides an accessible approach to decrease $P_f$ ($k_f<0$) as close to $P^{(1)}_f$ as possible while further decreasing $P_s$ ($k_s<0$) by using the present optimization algorithm.
Figure \ref{fig2}(b) shows the final optimized populations in the three states as a function of $A^2(\omega_0)$.
A linear response of $P_f$ with respect to $A^2(\omega_0)$ is restored while efficiently suppressing nonlinear optical transitions to the state $|s\rangle$. As a result, a linear superposition $\alpha|g\rangle+\beta|f\rangle$ is obtained, where the coefficients $\alpha$ and $\beta$ are complex numbers satisfying $|\alpha|^2+|\beta|^2=1$. In the field of quantum computing, this superposition state corresponds to  a qubit.  \\ \indent
The optimized spectral phases at different values of  $A^2(\omega_0)\leq 1.0$ are plotted in Fig. \ref{fig3}.
It is observed that the optimized spectral phases are mainly modulated around the
two fundamental frequencies $\omega_0$ and
$\omega_0-\omega_{sg} = 0.98\omega_0$, leading to  a substantial reduction   of  multi-photon (e.g., resonance  Raman) transitions to the state $|s\rangle$.
To gain insight into the effect of the optimized spectral phase on the dynamics of the final
state, Fig. \ref{fig4} shows a comparison  of $P^{(1)}_f(t)$ and
$P_f(t)$ with constant  and optimized  spectral phases at three different values of $A^2(\omega_0)$. For constant spectral phase, the pulse smoothly
transfers the population to the final state $|f\rangle$, where the differences
between the first-order perturbation simulations and the exact
solution to the Schr\"{o}dinger equation
imply that higher-order perturbations and therefore nonlinear
optical effects play a role.  The optimized spectral phases prolong
the pulse durations
from the femtosecond to the picosecond regime and almost restore the
behaviour of $P^{(1)}_f(t)$ under the first-order description,
especially at lower intensities (see Fig. \ref{fig4} a'),
where the high-order perturbation terms are rather weak.
It is noteworthy that first-order perturbation
theory correctly predicts all post-pulse transition probabilities between
0 and 1.
The slight differences of the transient dynamics between the first-order perturbation simulations and the exact calculations (see Figs. \ref{fig4} b' and c') can be attributed to the fact that the nonlinear optical transitions take place  during the laser-system interactions, but their contributions to the final absorption probability to the state
$|f\rangle$ are completely eliminated. Note that the transient dynamics
induced by the spectral phase optimization exhibit strong oscillations in the excited state population, which differ from the case of constant spectral phase.  Oscillatory dynamics of the excited state population in the perturbative
regime of the interaction have been observed experimentally by
linearly chirping the
spectral phase of a laser pulse \cite{prl:87:033001}.\\ \indent
\begin{figure}[!t]\centering
\resizebox{0.5\textwidth}{!}{%
  \includegraphics{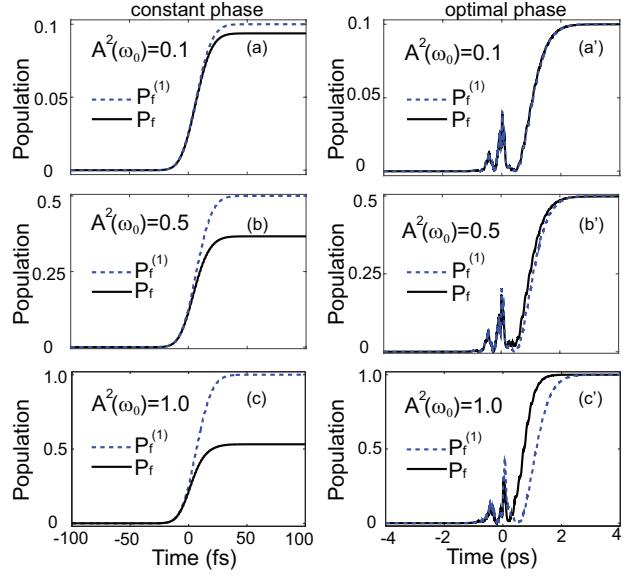}
} \caption{A comparison  of $P^{(1)}_f(t)$ (obtained by first-order perturbation calculation) and
$P_f(t)$ (the exact time-dependent Schr\"{o}dinger equation solution)  with constant (left panels) and
optimized (right panels) spectral phase pulses at (a) and (a') at $A^2(\omega_0)=0.1$, (b) and (b')
$A^2(\omega_0)=0.5$, and (c) and (c') $A^2(\omega_0)=1.0$. }\label{fig4}
\end{figure}
\begin{figure}[!t]\centering
\resizebox{0.5\textwidth}{!}{%
  \includegraphics{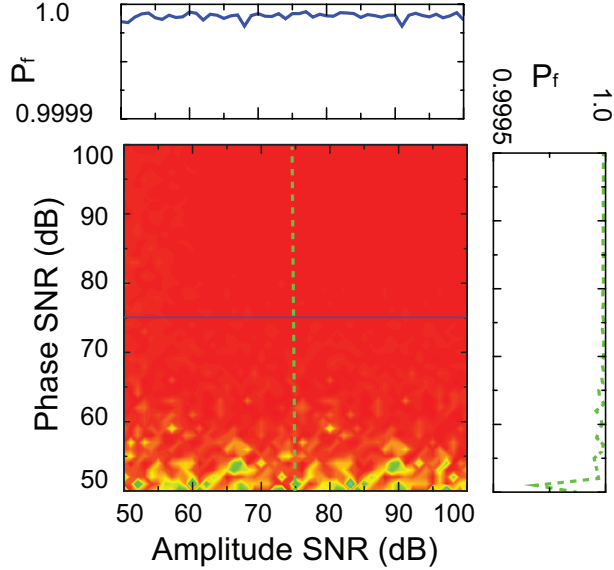}
} \caption{The robustness of the perfect  population transfer to  the final
state $|f\rangle$ (at $A^2(\omega_0)=1.0$) against the influence of spectral field noise. In this simulation,
a white Gaussian noise  of $50\sim100$ dB in signal-to-noise ratio (SNR) is added to the fixed spectral amplitude and the
optimized spectral phase, and then  the noised spectral field is transformed to obtain the temporal field of the laser pulse
for driving the evolution of the quantum system.}\label{fig5}
\end{figure}
We now examine the robustness of this scheme  against the influence of the control field noise, which has been identified as one of the key requirements in practical applications of quantum technology \cite{njp:11:105033,prl:111:050404}.  Due to various external or internal perturbations of laser sources, the temporal laser fields in laboratory  can be subject to stochastic
noise in either the time or frequency domain.  As an example, Fig. \ref{fig5} shows the final excited state population
variations versus the laser field fluctuations, where the fixed spectral amplitude $A(\omega)$ at $A^2(\omega_0) = 1.0$  and the optimized spectral phase
$\phi(\omega)$ are perturbed simultaneously with  white Gaussian noise  of $50\sim100$ dB in signal-to-noise ratio (SNR). A high efficiency of the population transfer  to the final state $|f\rangle$ can still be achieved  with an admissible error lower than $10^{-4}$ when the SNR is over 70 dB, which is possible using the current state-of-the-art laser techniques \cite{NPh:4:636}.
\\ \indent
In summary, we have presented an optical phase modulation scheme for coherent light
that can be utilized to completely eliminate nonlinear optical effects, leading to a linear absorption response from a low to a high energy level in
a multi-level quantum system.
The fundamental limit of single-photon absorption and therefore a linear superposition of two quantum states  was achieved  by transferring the optimal spectral phase of a broad bandwidth
ultrafast laser pulse onto the quantum wavefunction of the system.
To that end, a versatile spectral phase optimization algorithm was developed that can be used to monotonically approach multiple control objectives simultaneously.
This single-photon absorption limit was found  to be robust  with tolerable influence of spectral field noise.  These results suggest also an alternative approach to prepare  a  qubit in a multi-level quantum system. Since this multi-objective optimization algorithm  is general for maximizing the probability of single-photon transition while minimizing nonlinear optical transitions to multiple unwanted levels,   the key idea introduced  here could be extended to study  more complex atoms and molecules  as well as artificial quantum systems. This work can open a number of potential applications, including the manipulation of quantum wavefunctions, the extraction of a single photon from ultrafast laser pulses, and the storage of light information into quantum systems.

\begin{acknowledgements}
\indent D.D. and I.R.P. acknowledge partial  supports  by the Australian Research Council under Grant Nos. DP130101658 and  FL110100020. C.C.S acknowledges the financial support by the Vice-Chancellor's Postdoctoral Research Fellowship of University of New South Wales, Australia, and  the hospitality provided by the Technical University of Denmark during his visit in February, 2016.
\end{acknowledgements}

\end{document}